\documentclass[structabstract]{aa}
\usepackage{txfonts,graphicx,natbib}
\bibpunct{(}{)}{;}{a}{}{,}

\begin{document}

\title{AKARI near-infrared spectroscopy of the aromatic and aliphatic hydrocarbon emission features in the galactic superwind of M~82 }
\titlerunning{Hydrocarbon emission in the galactic superwind of M~82}

\author{M. Yamagishi\inst{1} \and H. Kaneda\inst{1} \and D. Ishihara\inst{1} \and T. Kondo\inst{1} \and T. Onaka\inst{2} \and T. Suzuki\inst{3} \and Y. C. Minh\inst{4}}


\institute{Graduate School of Science, Nagoya University, Furo-cho, Chikusa-ku, Nagoya 464-8602, Japan \and Graduate School of Science, The University of Tokyo, 7-3-1 Hongo, Bunkyo-ku, Tokyo 113-0033, Japan \and Institute of Space and Astronautical Science, Japan Aerospace Exploration Agency, 3-1-1 Yoshinodai, Chuo-ku, Sagamihara 252-5210, Japan \and Korea Astronomy and Space Science Institute, 776 Daedeok-daero, Yuseong, Daejeon 305-348, Korea}

\date{Received *** / Accepted ***}

\abstract {}
{We investigate the properties of hydrocarbon grains in the galactic superwind of M~82.}
{With AKARI, we performed near-infrared (2.5 -- 4.5 $\mathrm{\mu m}$) spectroscopic observations of 34 regions in M~82 including its northern and southern halos.}
{Many of the spectra show strong emission at 3.3 $\mathrm{\mu m}$ due to polycyclic aromatic hydrocarbons (PAHs) and relatively weak features at 3.4 -- 3.6 $\mathrm{\mu m}$ due to aliphatic hydrocarbons. In particular, we clearly detect the PAH 3.3 $\mathrm{\mu m}$ emission and the 3.4 -- 3.6 $\mathrm{\mu m}$ features in halo regions, which are located at a distance of 2 kpc away from the galactic center. We find that the ratios of the 3.4 -- 3.6 $\mathrm{\mu m}$ features to the 3.3 $\mathrm{\mu m}$ feature intensity significantly increase with distance from the galactic center, while the ratios of the 3.3 $\mathrm{\mu m}$ feature to the AKARI 7 $\mathrm{\mu m}$ band intensity do not.}
{Our results clearly confirm the presence of small PAHs even in a harsh environment of the halo of M~82. The results also reveal that the aliphatic hydrocarbons emitting the 3.4 -- 3.6 $\mathrm{\mu m}$ features are unusually abundant in the halo, suggesting that small carbonaceous grains are produced by shattering of larger grains in the galactic superwind.}


\maketitle 


\section{Introduction}

M~82 is a nearby edge-on starburst galaxy in the M~81--M~82 group.
Because of its proximity (3.53 Mpc, 1\arcsec = 17 pc ; \citealt{karachentsev02}), M~82 has been observed over a wide range of wavelengths (e.g., X-ray: \citealt{strickland04}; UV: \citealt{hoopes05}; optical: \citealt{oyama02} ; infrared: \citealt{engelbracht06}, \citealt{kaneda10}, \citealt{roussel10}; radio: \citealt{yun94}, \citealt{walter02}).
The starburst activity in M~82 was presumably triggered by close encounter with M~81 about 100 Myr ago (\citealt{yun93}).
As a result of the starburst activity, M~82 shows prominent galactic superwind, pushing out a significant amount of the interstellar medium from the galactic disk.
The outflows are observed in ionized gas (e.g., H$\alpha$: \citealt{oyama02}; X-ray: \citealt{strickland04}) with filamentary structures, which entrains neutral gas (e.g., CO: \citealt{walter02}; $\mathrm{H_2}$: \citealt{veilleux09}) and dust (e.g., \citealt{alton99}; \citealt{engelbracht06}; \citealt{leeuw09}; \citealt{kaneda10}; \citealt{roussel10}; \citealt{yoshida11}).

Polycyclic aromatic hydrocarbons (PAHs) show prominent emission features in the mid-infrared (mid-IR), which are important to characterize the interstellar medium especially in star-forming galaxies (e.g., \citealt{lu03}; \citealt{smith07}).
With Spitzer, \citet{engelbracht06} showed that the emission of PAH extends to 6 kpc from the galactic plane.
With AKARI, \citet{kaneda10} found a tight correlation between the PAH and the H$\alpha$ emission in the halo of the galaxy.
Since the plasma responsible for the H$\alpha$ emission shows outflows with the radial velocities of $\sim$ 600 $\mathrm{km~s^{-1}}$ (\citealt{shopbell98}), the PAHs are also very likely to be flowing out of the galaxy.
PAHs are, however, expected to be destroyed quite easily in X-ray hot plasma with temperature of $\gtrsim$ $10^6$ K, since PAHs are small ($\sim$ 6$\AA$; \citealt{tielens08}) particles with 2-dimensional structures.
For example, \citet{micelotta10b} calculated that the timescale of the PAH destruction by collisions with electrons in the X-ray superwind of M~82 is less than a thousand years.
Therefore it is unlikely that the PAHs coexist with the X-ray hot plasma in the halo.
PAHs are probably in dense molecular clouds protected against the destruction (\citealt{veilleux09}), or in a much cooler (H$\alpha$) plasma phase even if they coexist with the plasma (\citealt{kaneda10}).

In this paper, we present the near-IR (2.5 -- 4.5 $\mathrm{\mu m}$) spectra of M~82 obtained with the Infrared Camera (IRC; \citealt{onaka07}) on board AKARI (\citealt{murakami07}).
The spectral range includes the PAH 3.3 $\mathrm{\mu m}$ emission feature and the 3.4 -- 3.6 $\mathrm{\mu m}$ features.
We clearly detect these features from the halo of M~82.
Both features are attributed to C-H vibration modes of carbonaceous grains; the former is due to aromatic hydrocarbons, while the latter is aliphatic hydrocarbons (\citealt{duley81}).
Based upon these features, we discuss the properties of carbonaceous grains including PAHs in the galactic superwind of M~82.

\section{Observations and data reductions}
The IRC spectroscopic observations of M~82 were carried out in October 2007 and August 2008.
A summary of the observations is listed in table 1.
All the data were taken from the AKARI archives; the first three observational data are publicly available on the official site\footnote{http://darts.jaxa.jp/ir/akari/}, while the last two are not because they were taken by using the instrumental calibration time.
To obtain the near-IR (2.5 -- 4.5 $\mathrm{\mu m}$) spectra, the grism spectroscopic mode was used in the observations.
Although the data covered the wavelength range of 2.5 -- 5.0 $\mathrm{\mu m}$, we did not use the range of 4.5 -- 5.0 $\mathrm{\mu m}$, since the signal-to-noise ratios were relatively low at these wavelengths.
We extracted spectral data from the two slits Ns and Nh with the sizes of 5\arcsec $\times$ 48\arcsec and 3\arcsec $\times$ 60\arcsec, respectively (\citealt{oyama07}).
They have different spectral resolutions for diffuse sources (Ns: R $\sim$ 100, Nh: R $\sim$ 150) due to the difference in the slit width.
Depending on the signal intensity, we divided the slit aperture area into two, three, or six and integrated the sub-slit data to derive a spectrum.
As a result, we obtain 34 spectra from the whole galaxy.
Figure 1 shows the positions of the slit sub-apertures on the AKARI 7 $\mathrm{\mu m}$ band map, from which we created spectra.
The AKARI 7 $\mathrm{\mu m}$ band (Fig. 1) covers the wavelength range of 5.9 -- 8.4 $\mathrm{\mu m}$ (\citealt{onaka07}), which includes the PAH 6.2 and 7.7 $\mathrm{\mu m}$ features.
Hereafter, we define the center, disk, and halo regions by the areas demarcated by the 7 $\mathrm{\mu m}$ surface brightness levels of 40 \% and 0.32 \% of the peak, which correspond to the third and the 10th contour from the top in Fig. 1.
According to this definition, sub-aperture regions A2-4 belong to the center, regions A1, A5-6, B1-6, C1-6, D1-3, and E1-2 to the disk, and regions E3, F1-2, G1-2, H1-2, I1-2, and J1-2 to the halo.
The basic spectral analyses were performed by using the official pipeline prepared for reducing phase 3 data (IRC Spectroscopy Toolkit for Phase 3 data Version 20110301).
In addition to the basic pipeline process, we applied the custom procedures to improve signal-to-noise ratios for each spectrum as described in \citet{yamagishi11}.
Finally, we applied smoothing with a boxcar kernel of $\sim$ 0.05 $\mathrm{\mu m}$ (R $\sim$ 60) to both Ns and Nh spectra to mitigate the effect of the difference in the slit width (i.e. resolution) between the Ns (R $\sim$ 100) and Nh (R $\sim$ 150) slits.

\begin{table}
\caption{Summary of the observations and sub-slit apertures used in the present study} 
\label{table:1} 
\centering 
\begin{tabular}{c c c c c} 
\hline\hline 
Obs. date   & Obs. ID & slit & sub-slit apertures & aperture size \\ 
\hline 
2008 Aug 21 & 3390001.1 & Nh & A1-6 & 3\arcsec $\times$ 7\farcs5 \\
            &         	& Ns & H1-2 & 5\arcsec $\times$ 18\arcsec \\
2008 Aug 21 & 3390002.1 & Nh & C1-6 & 3\arcsec $\times$ 7\farcs5 \\
            &         	& Ns & I1-2 & 5\arcsec $\times$ 18\arcsec \\
2008 Aug 21 & 3390003.1 & Nh & B1-6 & 3\arcsec $\times$ 7\farcs5 \\
            &         	& Ns & G1-2 & 5\arcsec $\times$ 18\arcsec \\
2007 Oct 23 & 5125401.1 & Nh & J1-2 & 3\arcsec $\times$ 22\farcs5 \\
            &         	& Ns & E1-3 & 5\arcsec $\times$ 12\arcsec \\
2007 Oct 22 & 5125405.1 & Nh & D1-3 & 3\arcsec $\times$ 15\arcsec \\
            &         	& Ns & F1-2 & 5\arcsec $\times$ 18\arcsec \\
\hline 
\end{tabular}
\end{table}

\begin{figure}
\centering
\resizebox{\hsize}{!}{\includegraphics{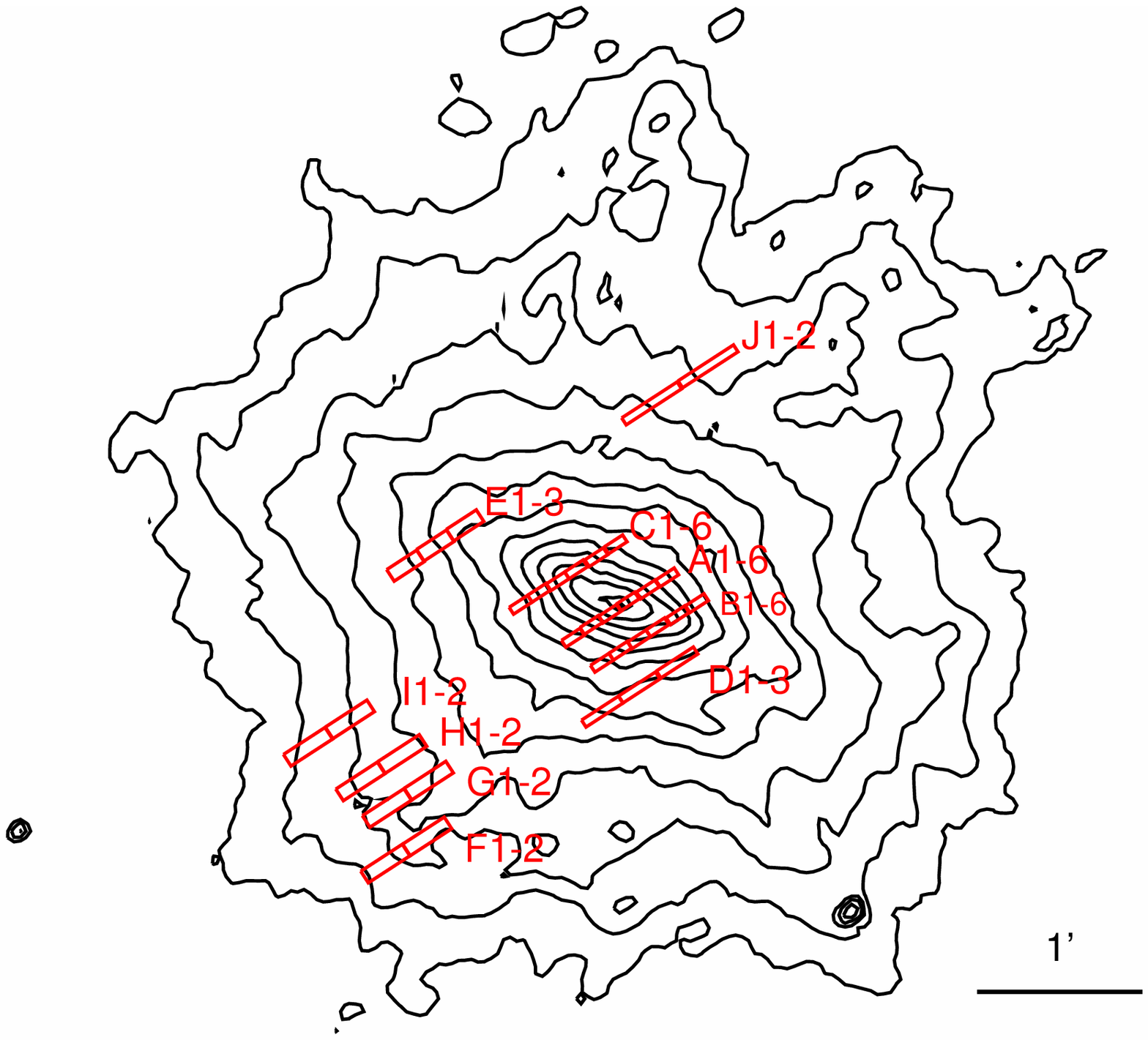}}
\caption{Positions of the sub-slit apertures used to derive the 34 spectra in Fig. 2, overlaid on the AKARI 7 $\mathrm{\mu m}$ band contour map (\citealt{kaneda10}). The north is up and the west to the right. The size of each box represents the area of each sub-slit aperture, which is listed in Table 1. The name of each region is indicated together, where the incremental order of the number is from the north to the south.}
\label{fig1}
\end{figure}

\section{Results}

Figure 2 shows the spectra obtained from various regions of M~82.
Many of the spectra show the aromatic hydrocarbon feature at 3.3 $\mathrm{\mu m}$ and the aliphatic features at 3.4 -- 3.6 $\mathrm{\mu m}$.
In particular, the spectrum in the halo of region F1 clearly shows the presence of the aromatic emission although the region is located at the distance so far as $\sim$ 2 kpc away from the galactic center.
In order to quantify the 3.3 -- 3.6 $\mathrm{\mu m}$ features, we approximate a linear baseline which is determined at the wavelength ranges of 3.0 -- 3.2 and 3.7 -- 3.9 $\mathrm{\mu m}$ on both sides of the features, as shown by the blue lines in Fig. 2.
Then, for the halo spectra, we calculate a standard deviation from a linear baseline at 3.8 -- 4.5 $\mathrm{\mu m}$ and adopt the 5$\sigma$ error in evaluating the significance of the 3.3 -- 3.6 $\mathrm{\mu m}$ excess emission.
For example, the spectrum in region F1 shows significance of 12$\sigma$ for the presence of the 3.3 -- 3.6 $\mathrm{\mu m}$ excess emission.
As a result, the excess emission is significantly detected from the spectra of all the regions except F2, H2, I1, and I2.

\begin{figure*}
\centering
\includegraphics[width=0.205\textwidth]{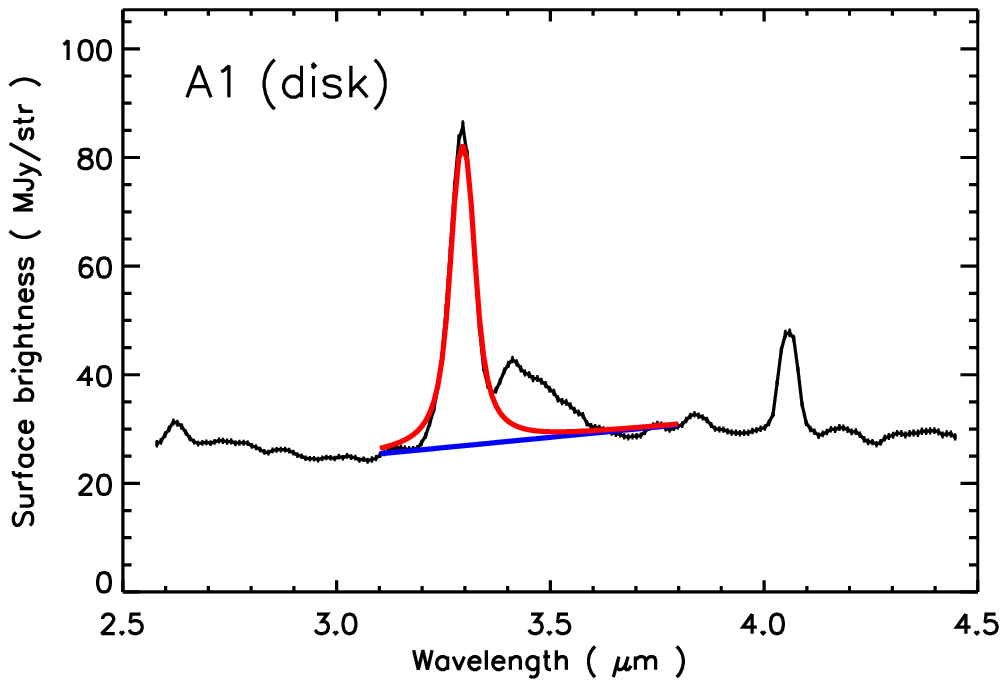}
\includegraphics[width=0.205\textwidth]{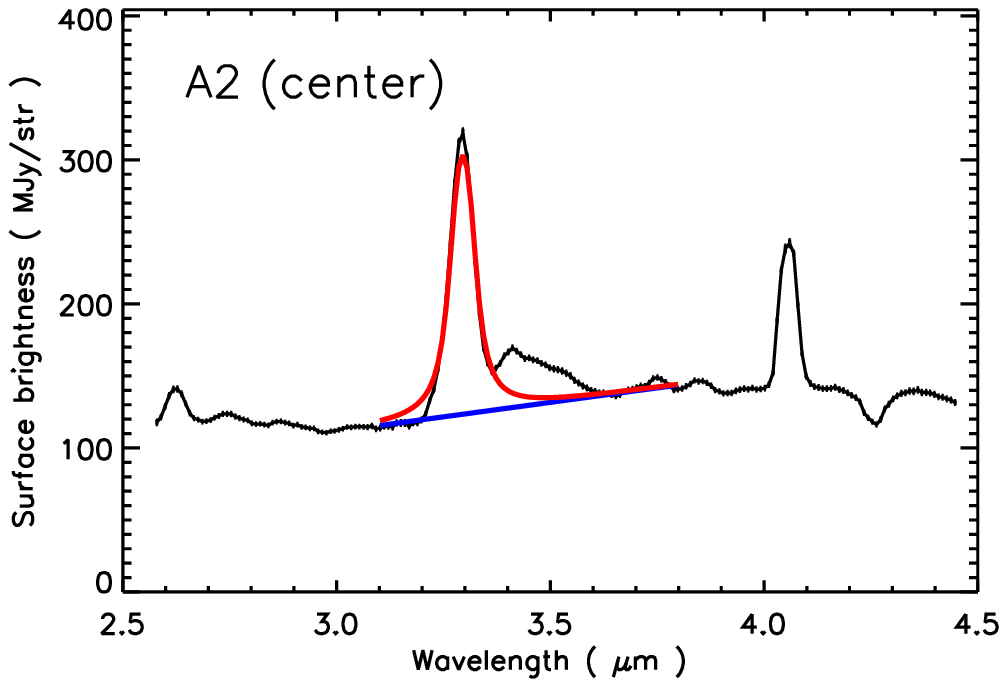}
\includegraphics[width=0.205\textwidth]{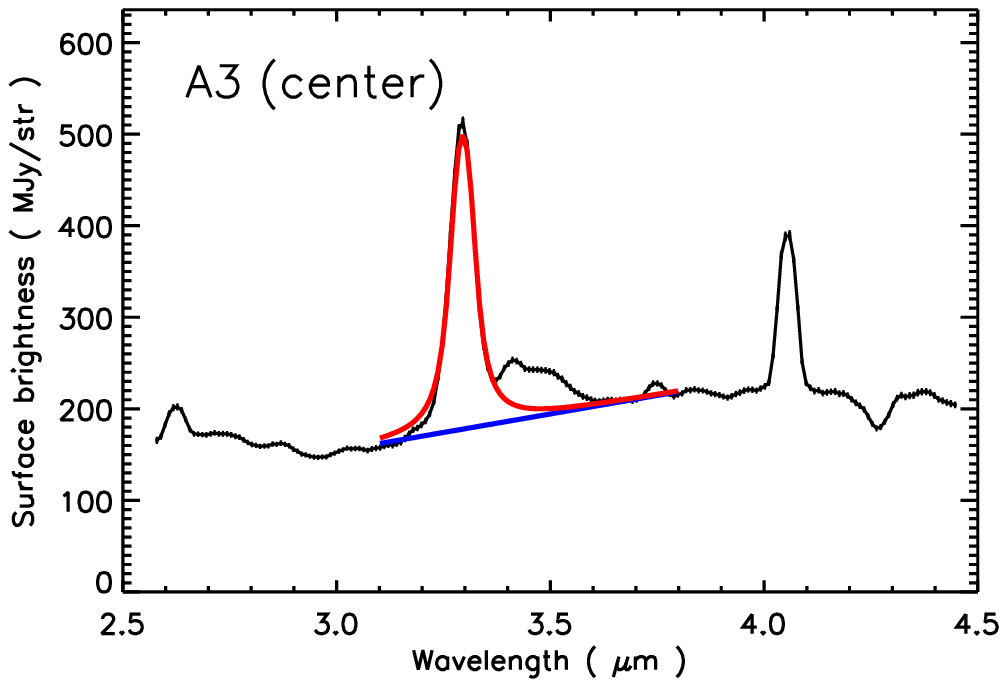}
\includegraphics[width=0.205\textwidth]{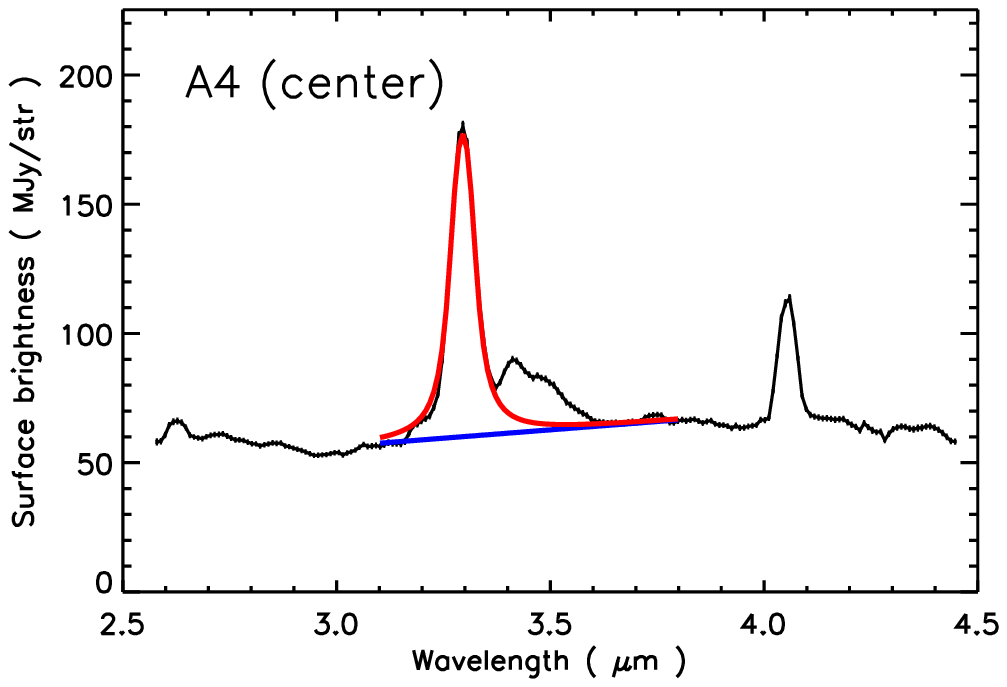}
\includegraphics[width=0.205\textwidth]{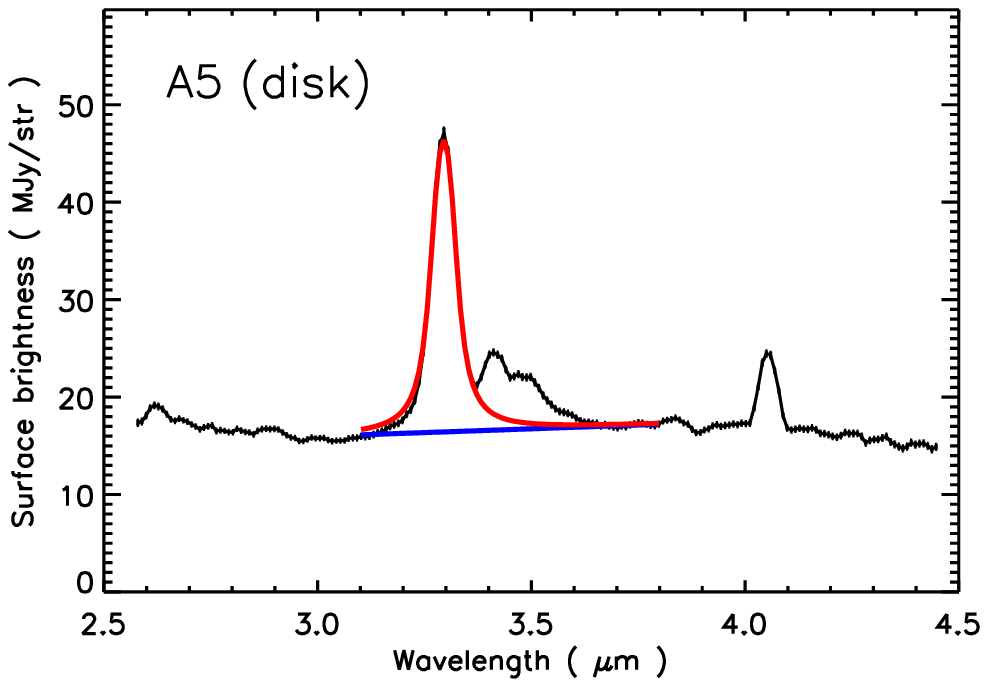}
\includegraphics[width=0.205\textwidth]{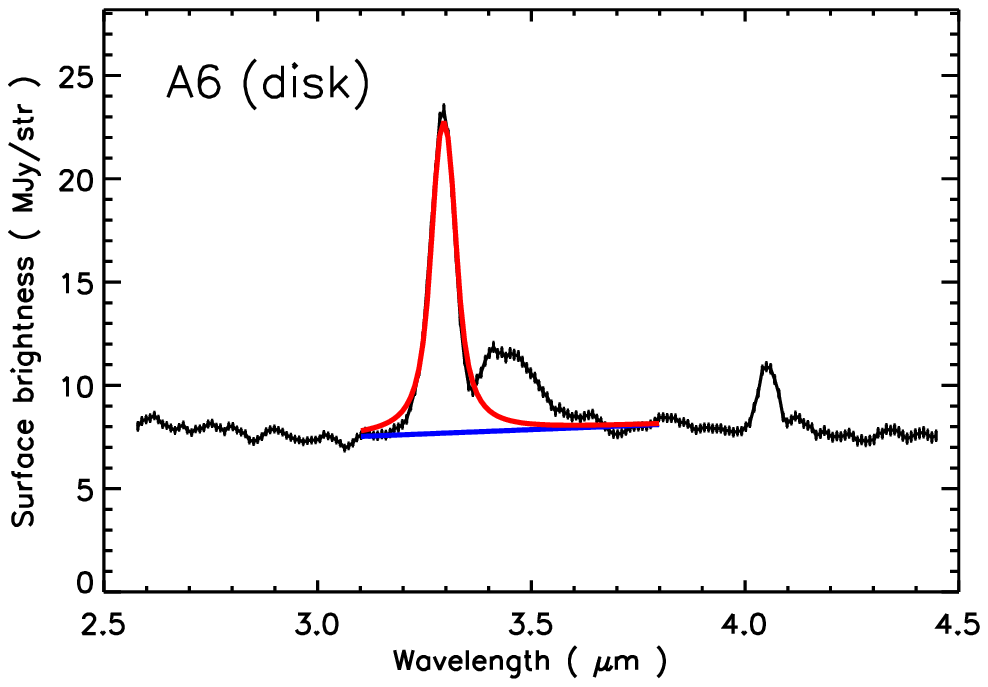}
\includegraphics[width=0.205\textwidth]{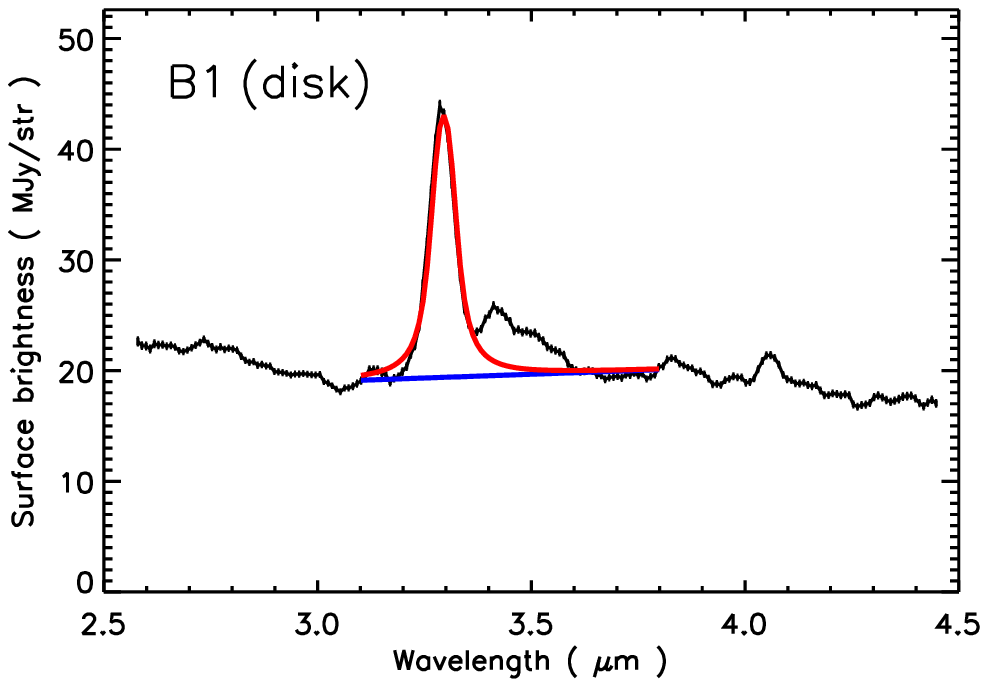}
\includegraphics[width=0.205\textwidth]{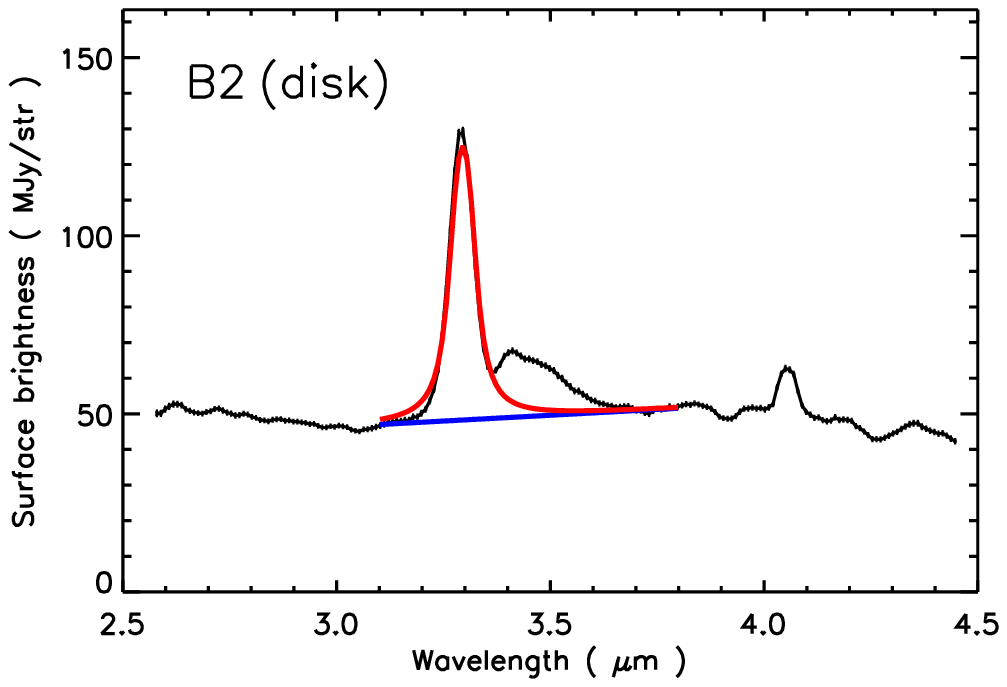}
\includegraphics[width=0.205\textwidth]{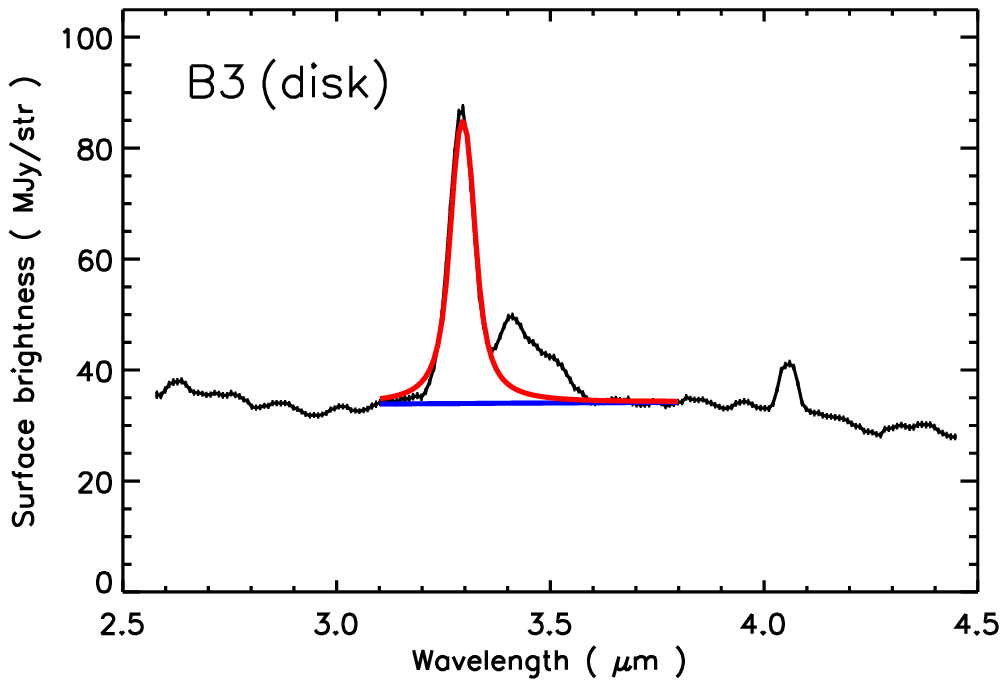}
\includegraphics[width=0.205\textwidth]{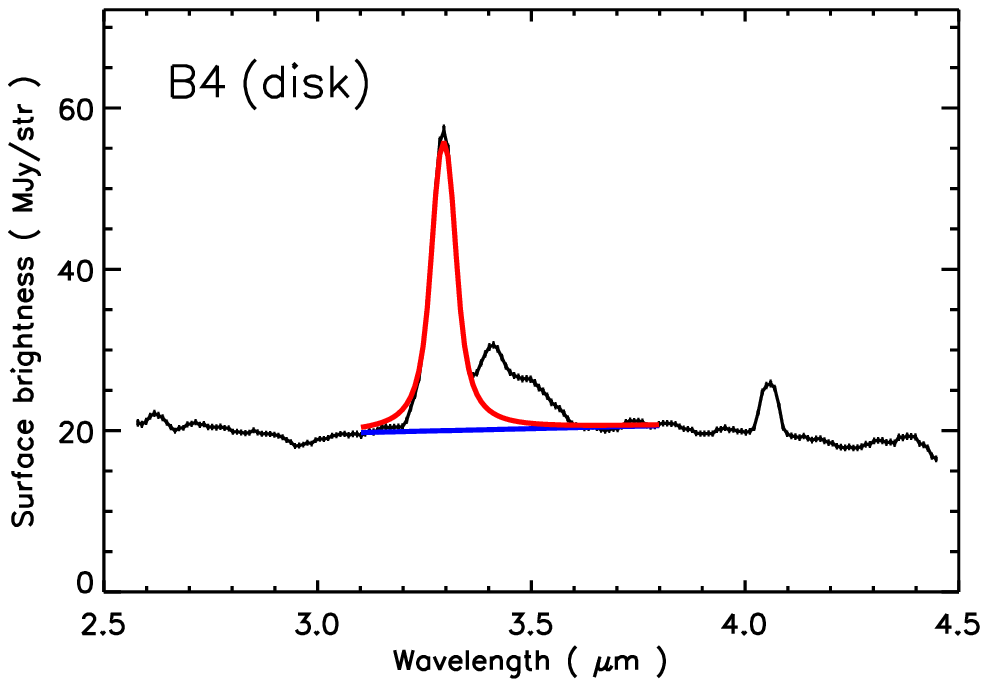}
\includegraphics[width=0.205\textwidth]{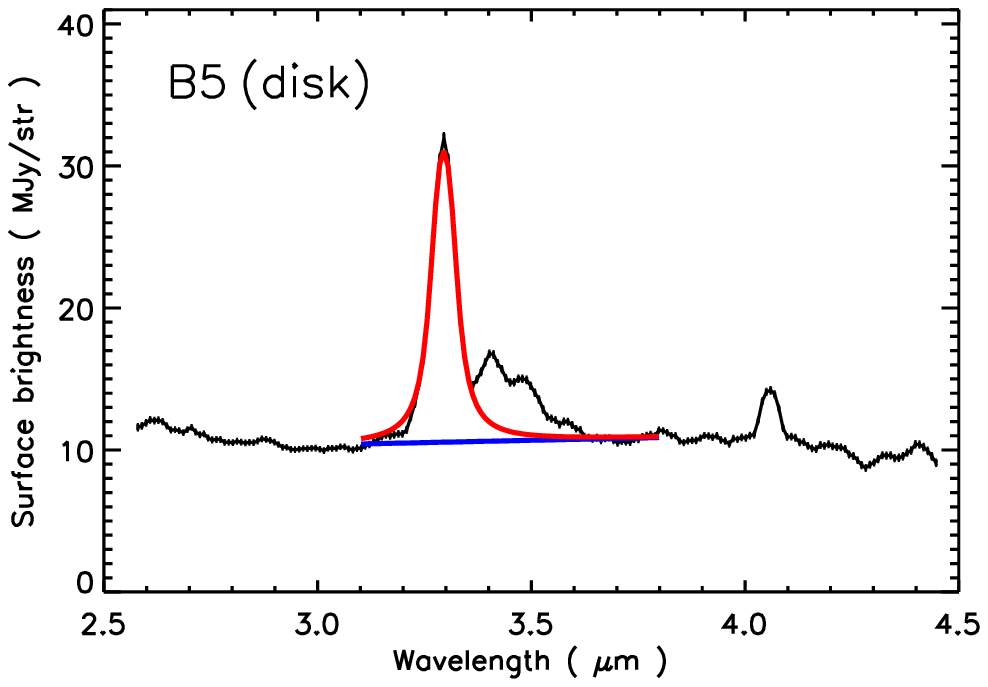}
\includegraphics[width=0.205\textwidth]{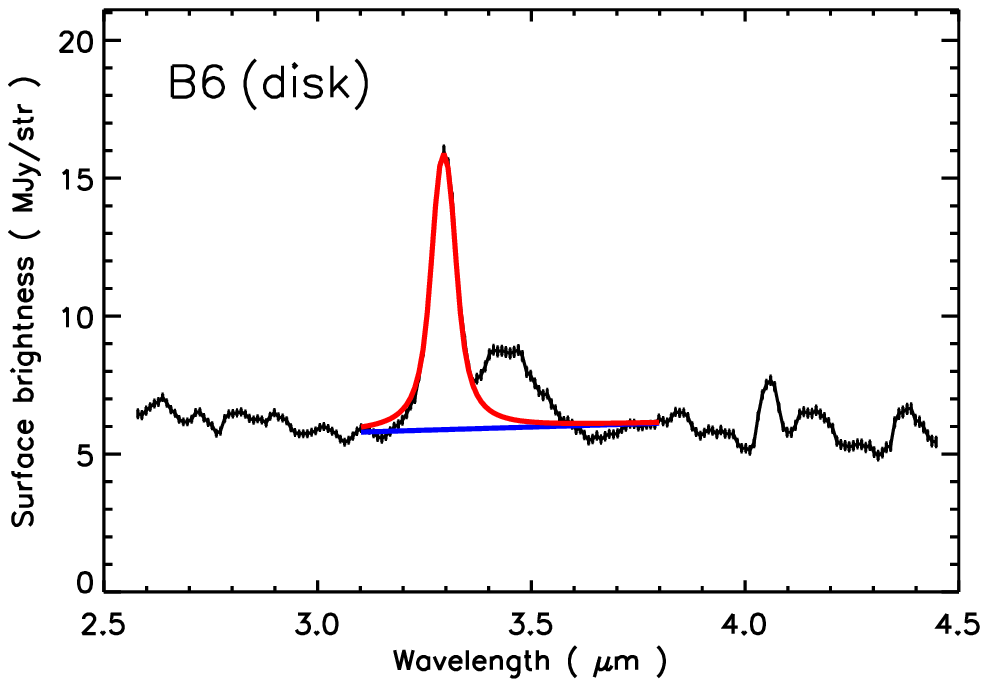}
\includegraphics[width=0.205\textwidth]{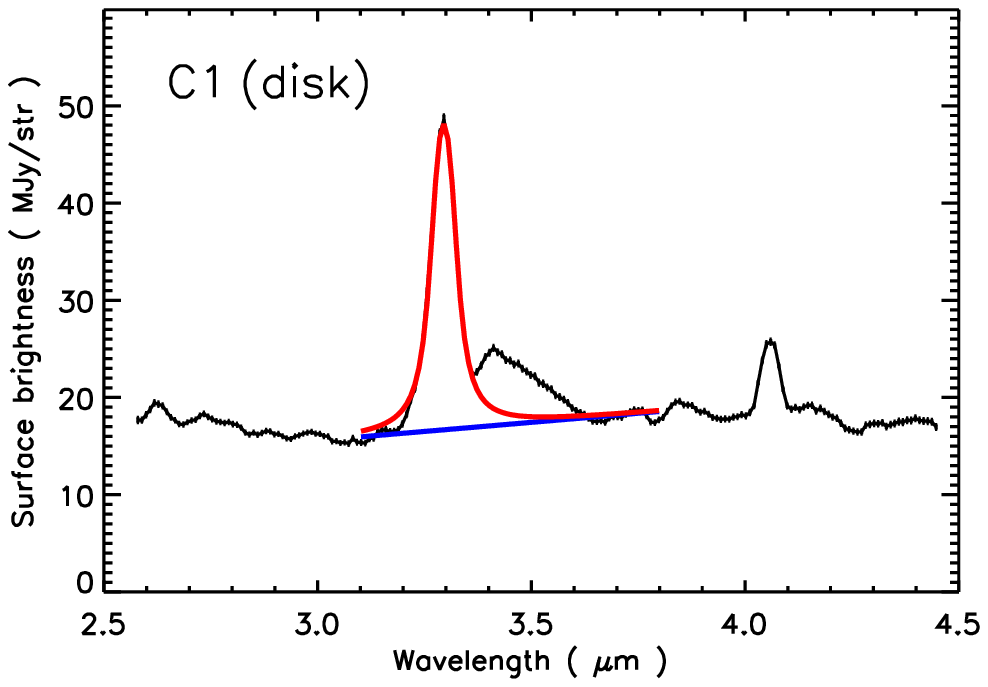}
\includegraphics[width=0.205\textwidth]{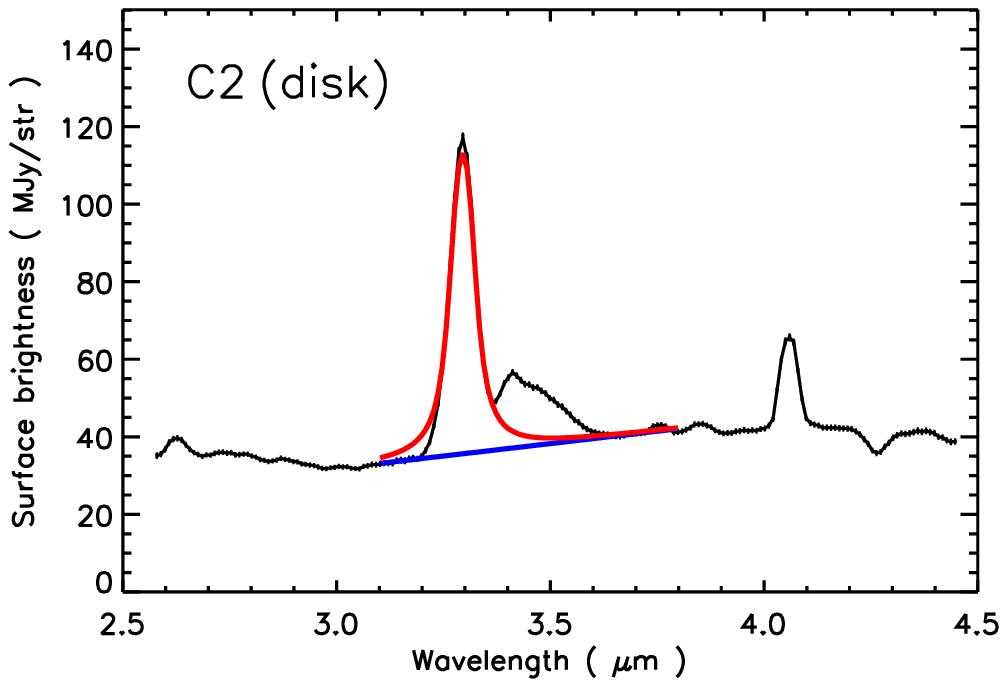}
\includegraphics[width=0.205\textwidth]{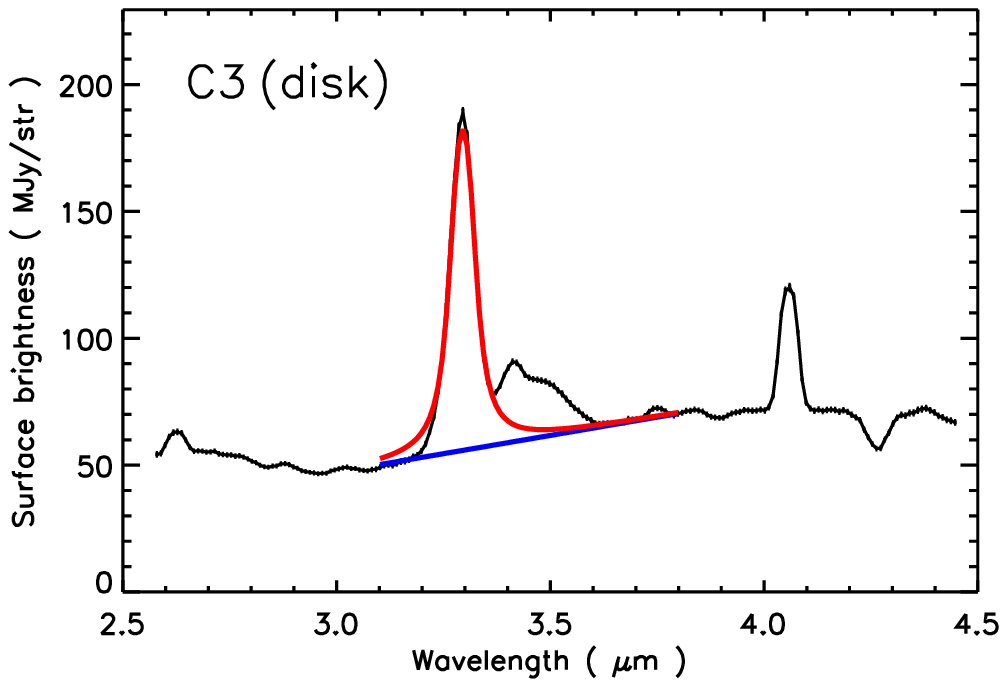}
\includegraphics[width=0.205\textwidth]{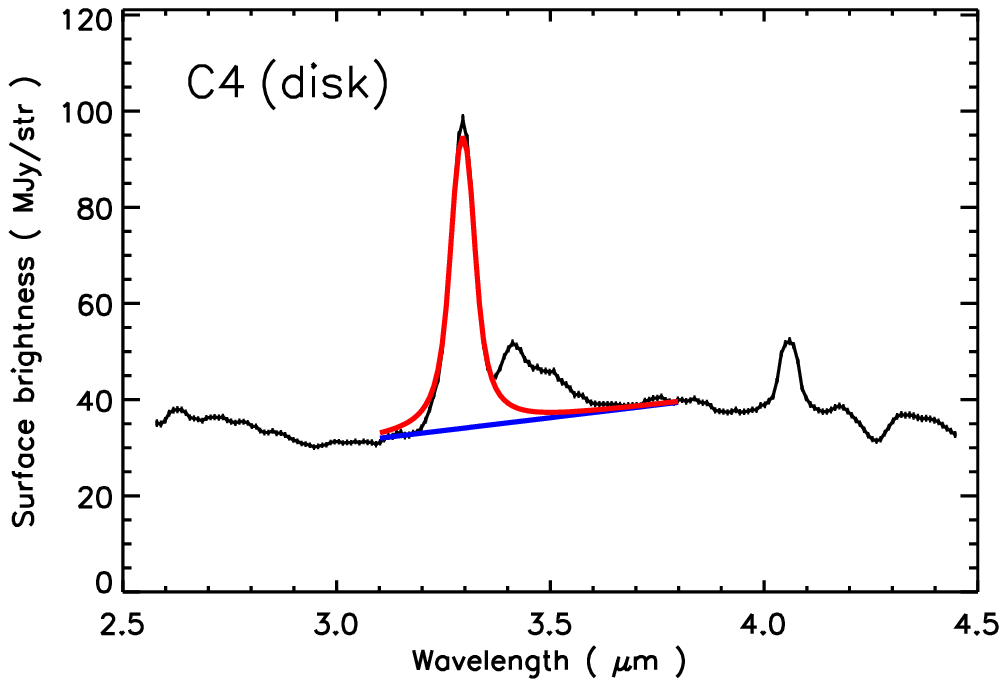}
\includegraphics[width=0.205\textwidth]{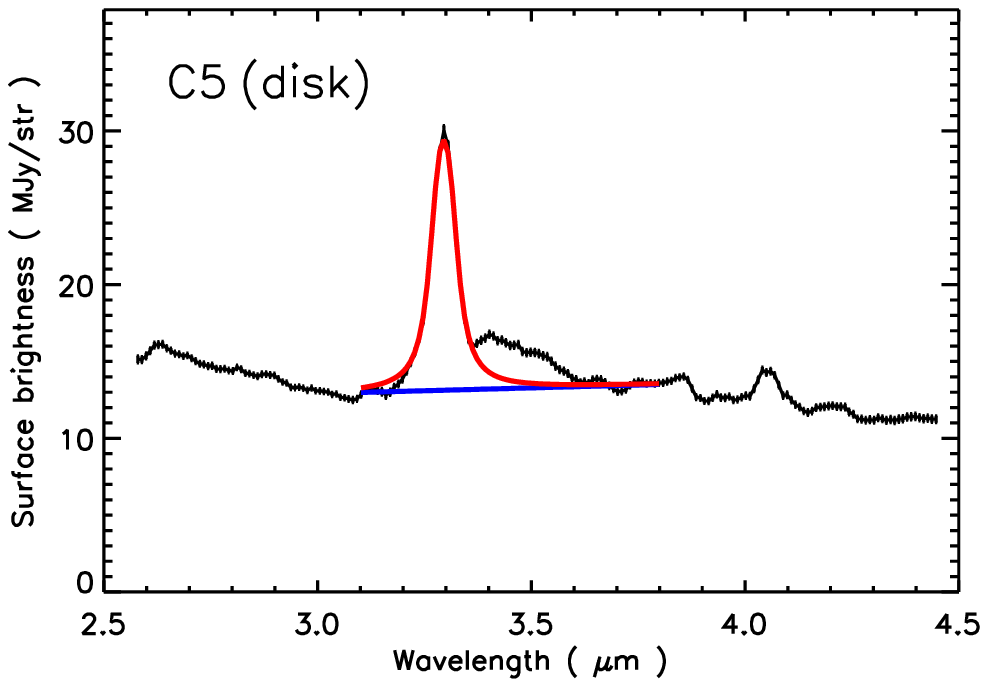}
\includegraphics[width=0.205\textwidth]{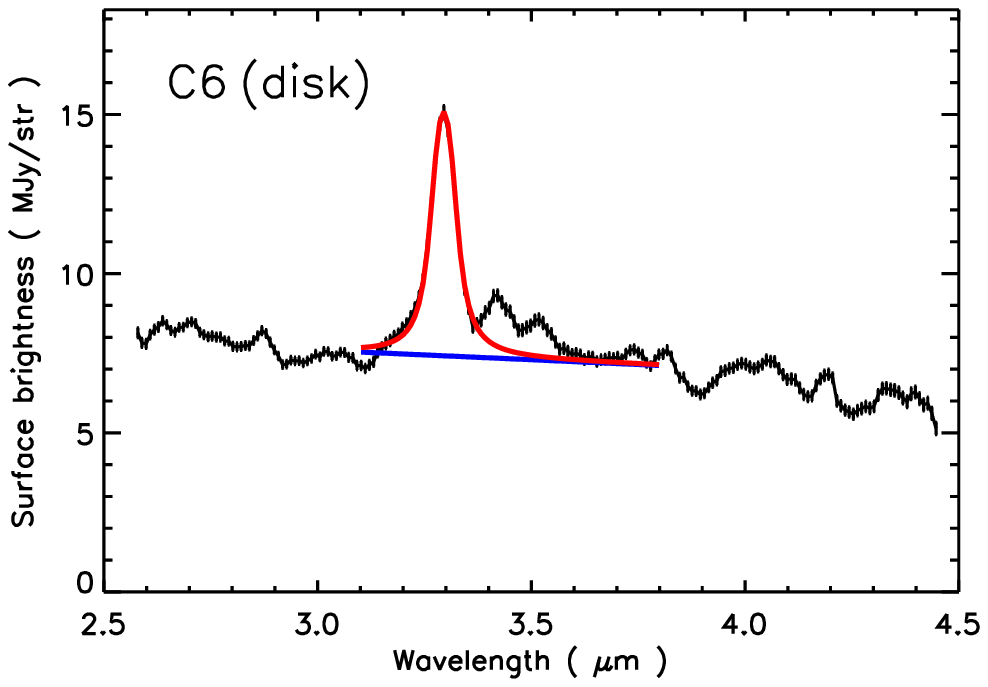}
\includegraphics[width=0.205\textwidth]{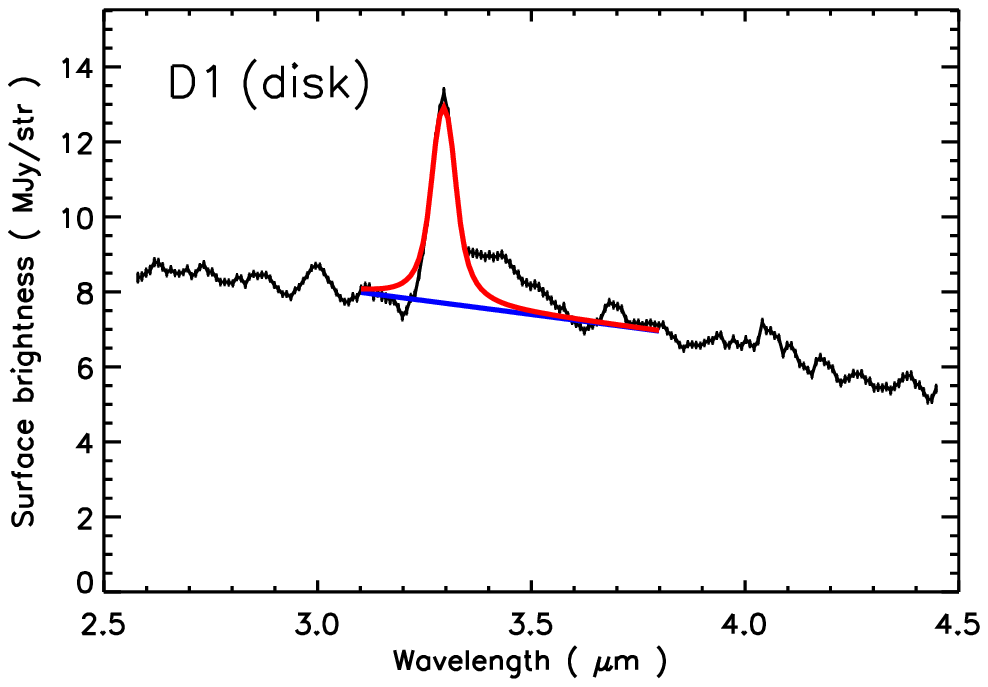}
\includegraphics[width=0.205\textwidth]{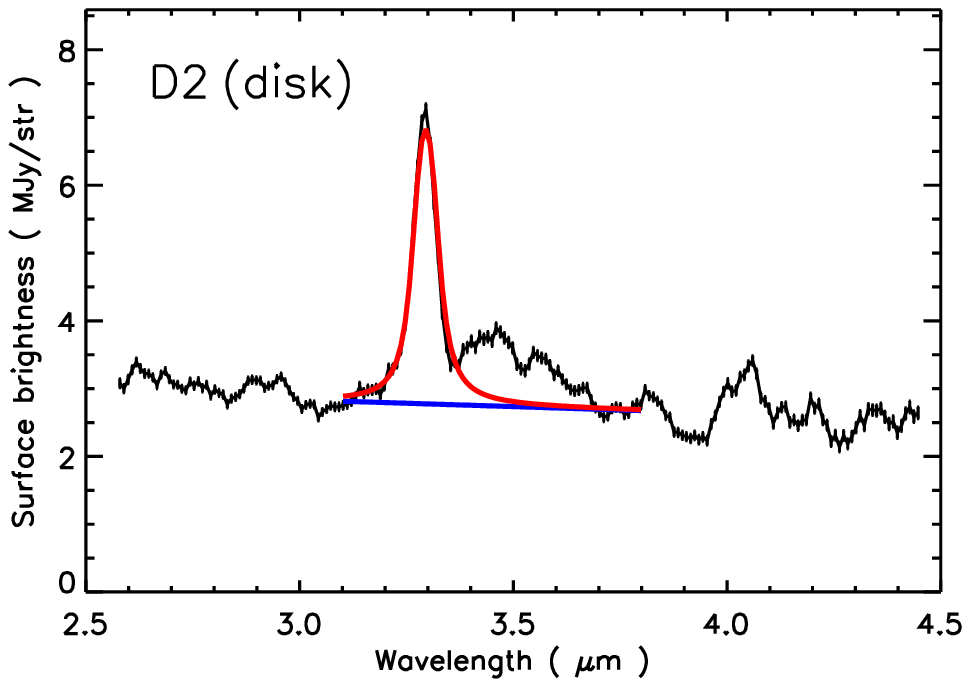}
\includegraphics[width=0.205\textwidth]{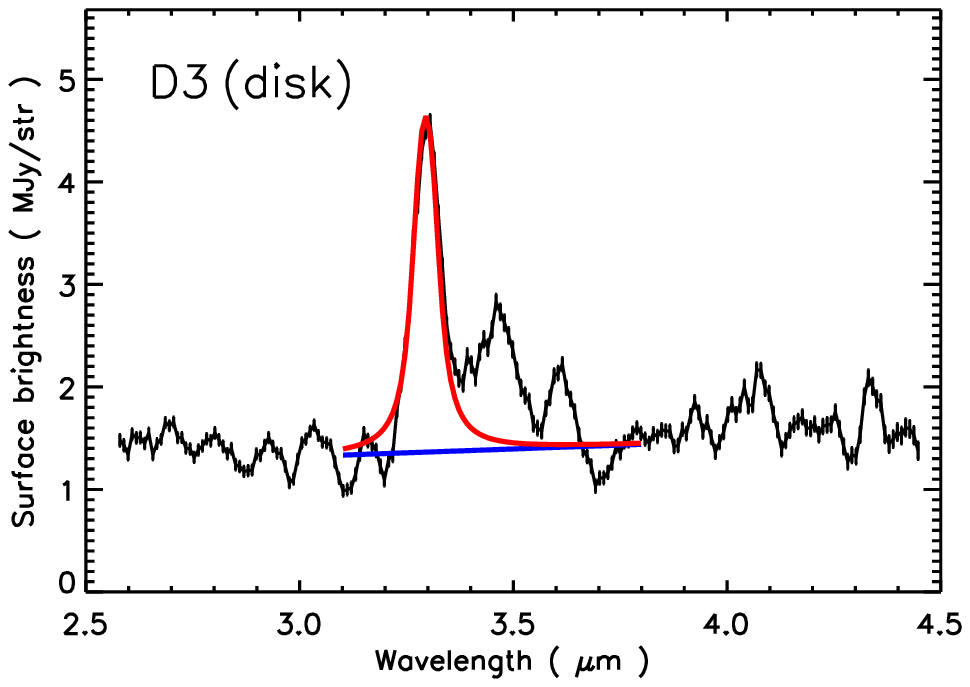}
\includegraphics[width=0.205\textwidth]{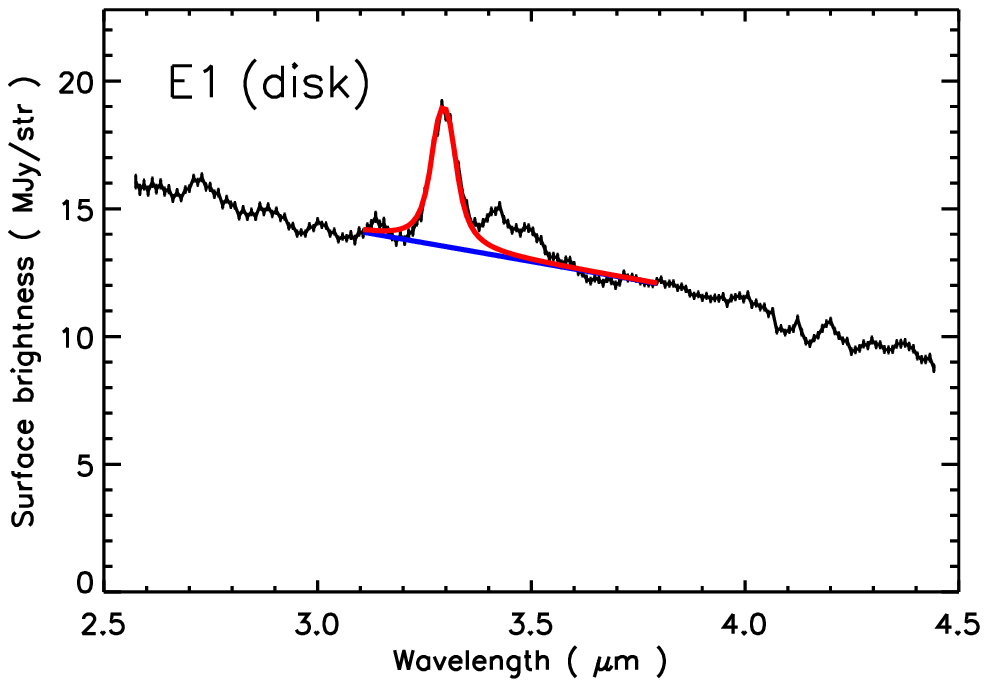}
\includegraphics[width=0.205\textwidth]{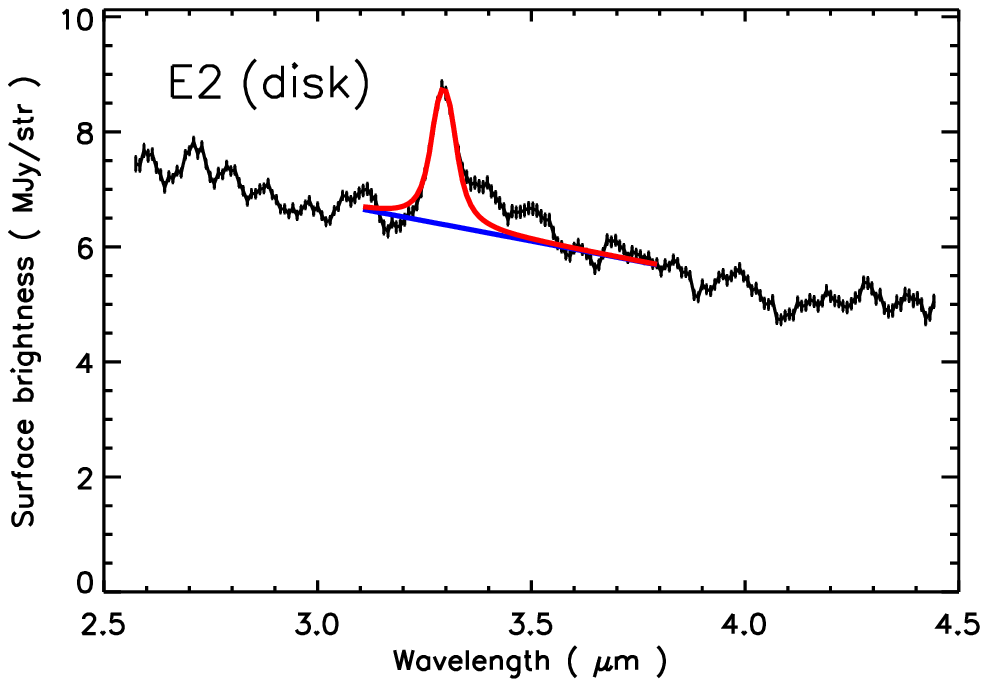}
\includegraphics[width=0.205\textwidth]{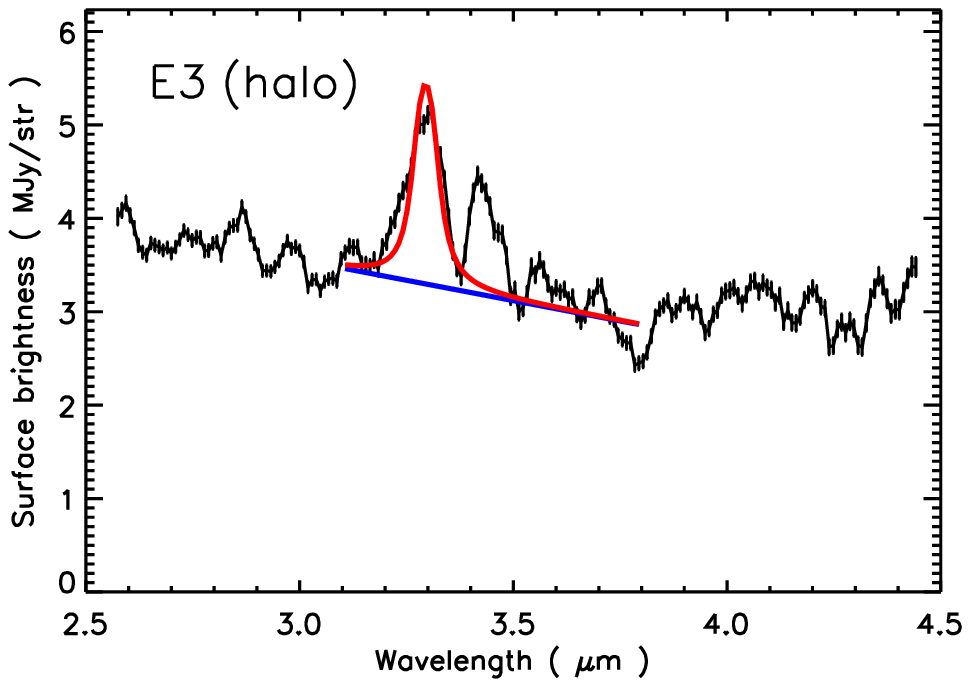}
\includegraphics[width=0.205\textwidth]{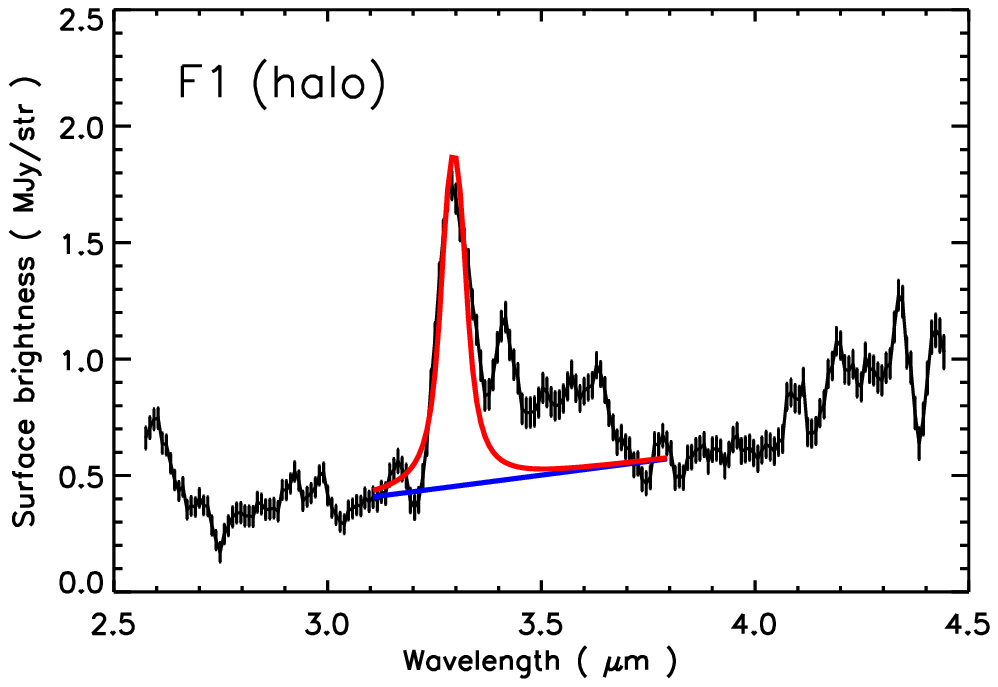}
\includegraphics[width=0.205\textwidth]{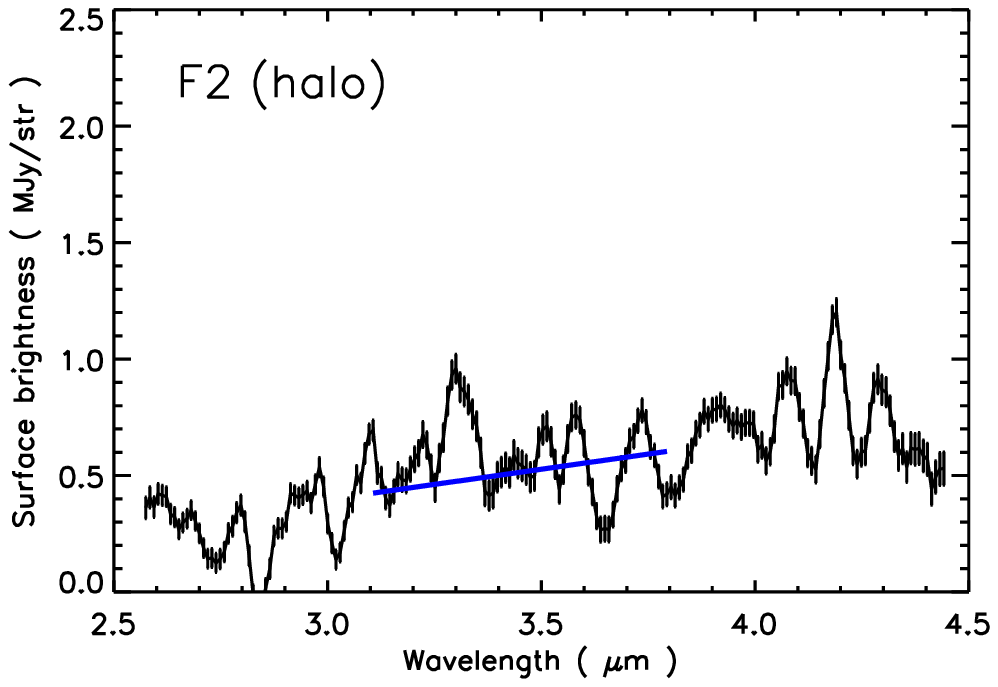}
\includegraphics[width=0.205\textwidth]{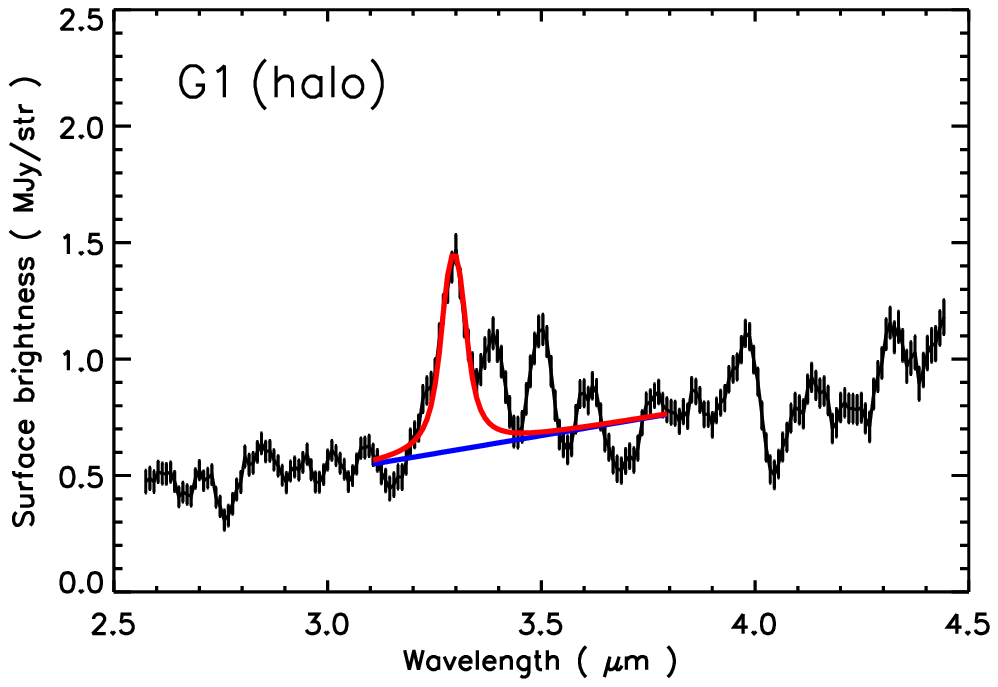}
\includegraphics[width=0.205\textwidth]{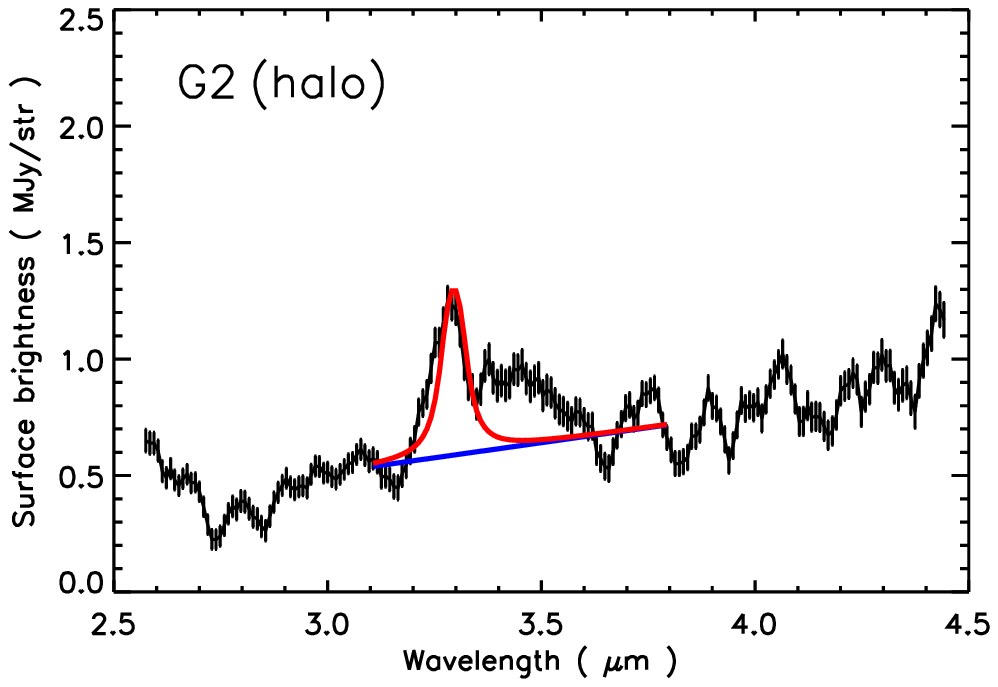}
\includegraphics[width=0.205\textwidth]{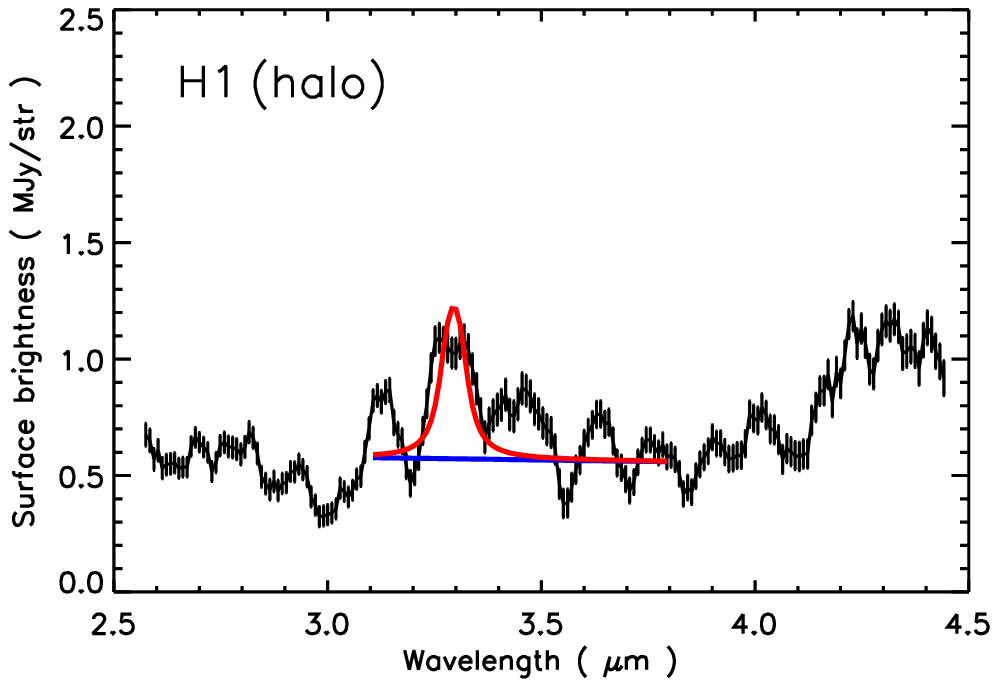}
\includegraphics[width=0.205\textwidth]{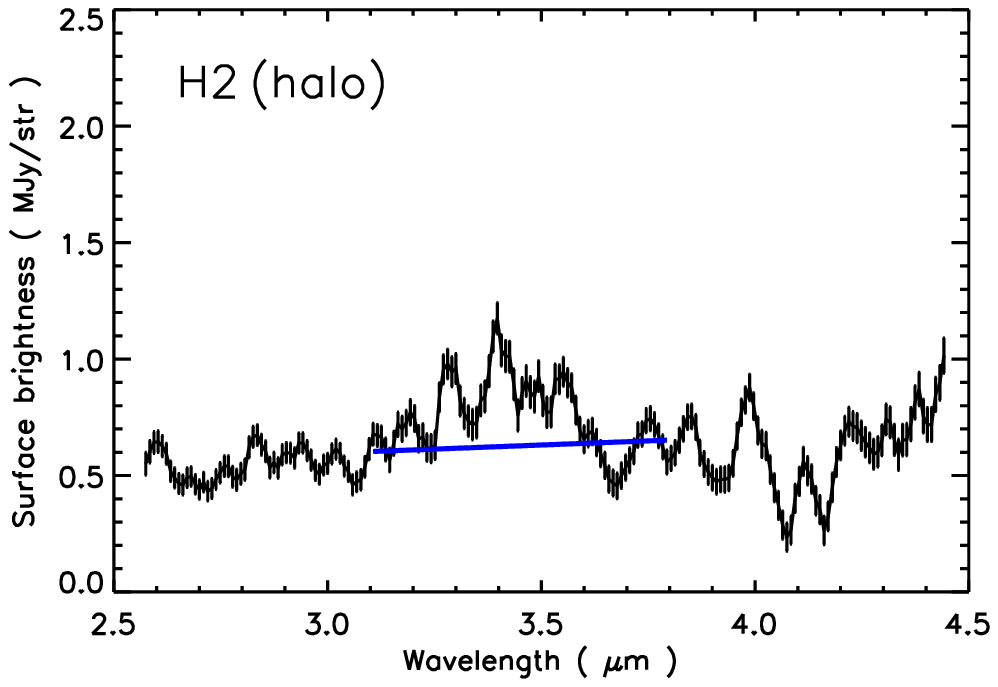}
\includegraphics[width=0.205\textwidth]{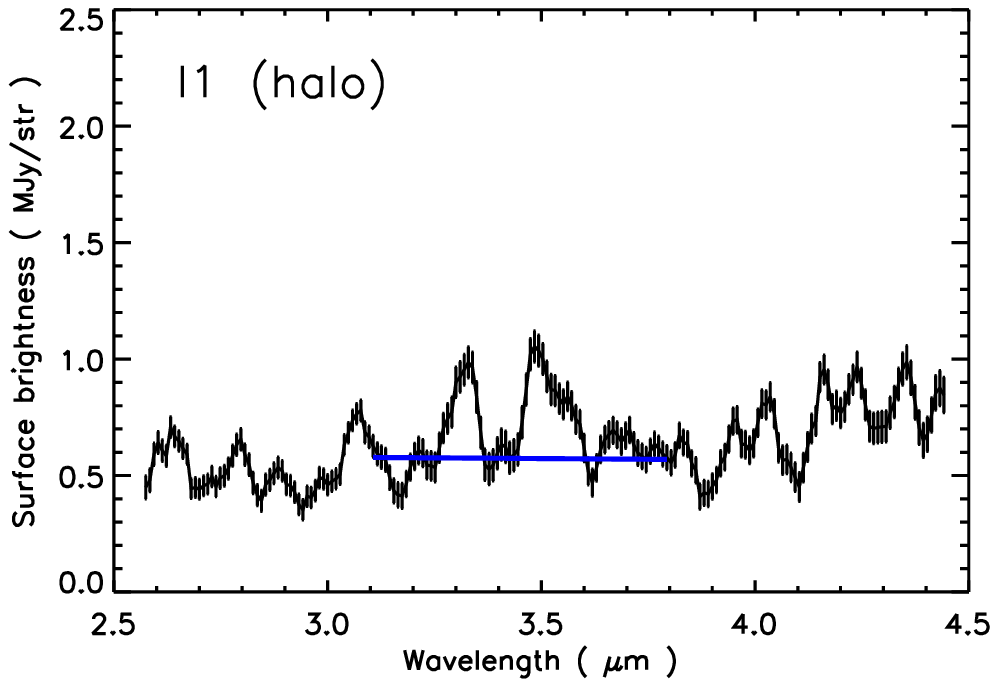}
\includegraphics[width=0.205\textwidth]{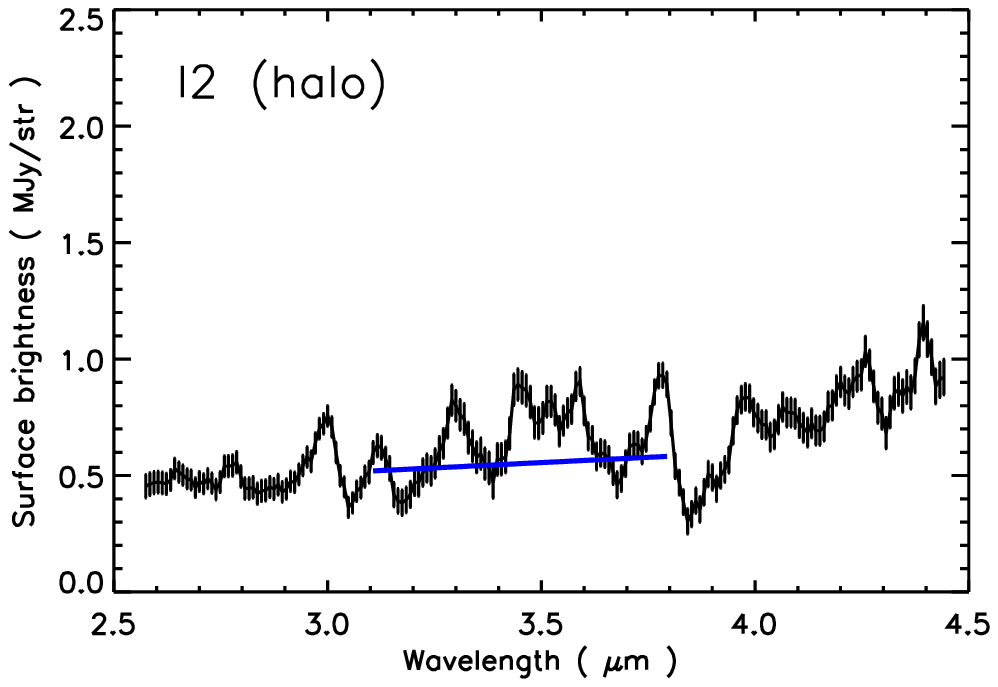}
\includegraphics[width=0.205\textwidth]{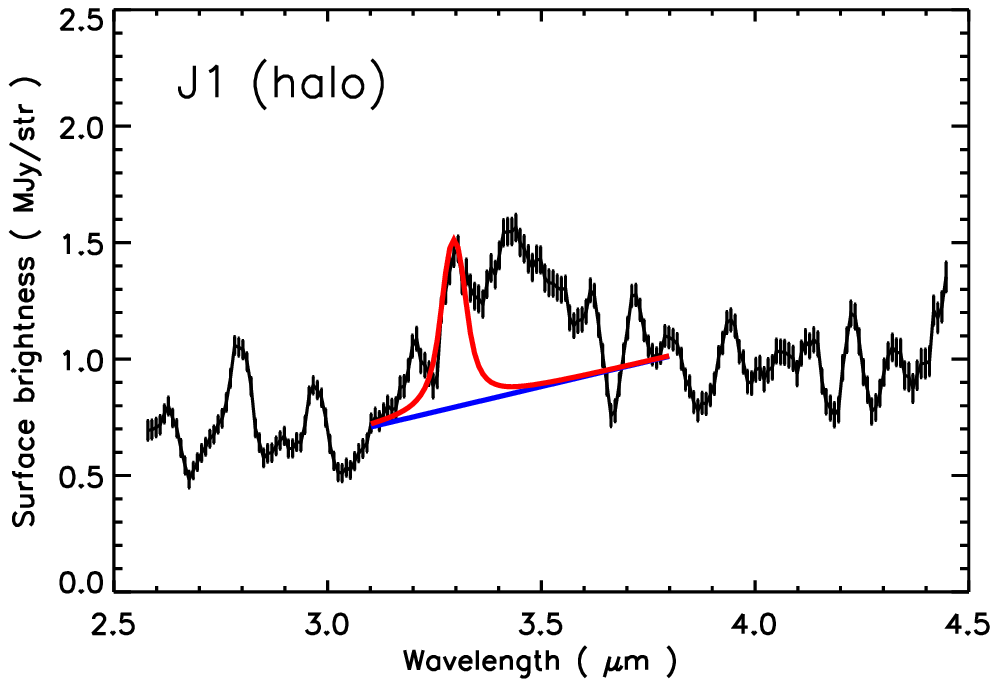}
\includegraphics[width=0.205\textwidth]{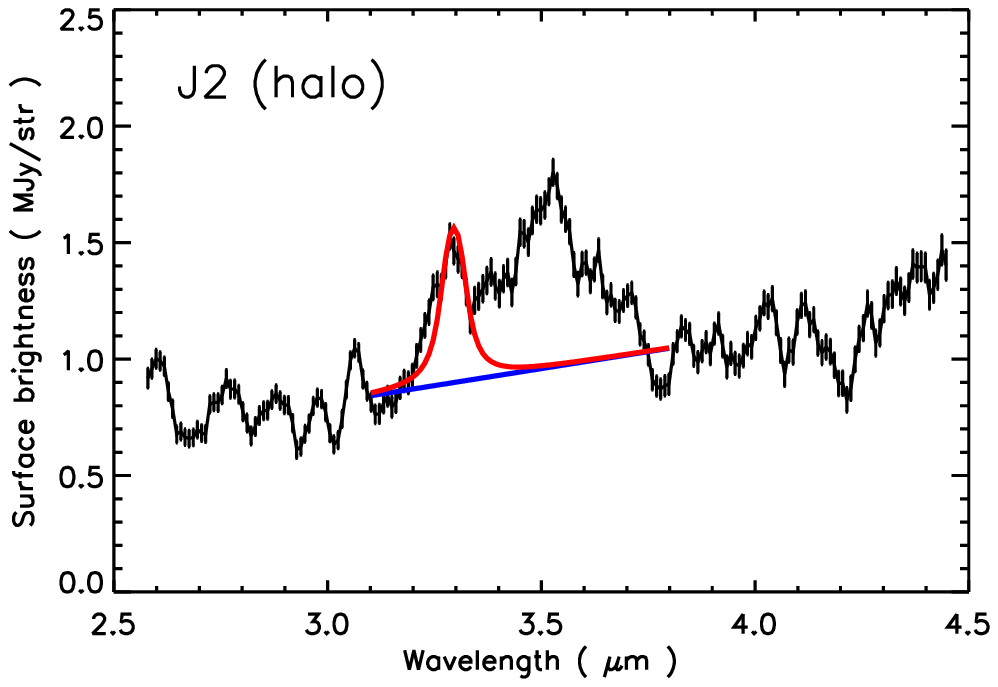}
\hspace{7.65cm}
\caption{AKARI 2.5 -- 4.5 $\mathrm{\mu m}$ spectra taken from the 34 regions in M~82, the positions of which are shown in Fig. 1. The blue linear baselines are determined at the wavelength ranges of 3.0 -- 3.2 $\mathrm{\mu m}$ and 3.7 -- 3.9 $\mathrm{\mu m}$. For the spectra with significant detection of the excess emission, the red curves indicate the best-fit Drude profiles representing the contribution of the aromatic 3.3 $\mathrm{\mu m}$ emission, which are determined at the wavelength range of 3.20 -- 3.35 $\mathrm{\mu m}$.}
\label{fig2}
\end{figure*}

The presence of the PAH emission extended in the halo was already reported by the mid-IR imaging observations (\citealt{engelbracht06}; \citealt{kaneda10}).
Spectroscopically, however, \citet{engelbracht06} confirmed the presence of only the PAH 17 $\mathrm{\mu m}$ features in the halo region.
Thus we for the first time show the presence of the aromatic 3.3 $\mathrm{\mu m}$ emission in the halo. Near the center region, hydrogen recombination lines (Br$\alpha$ at 4.05 $\mathrm{\mu m}$ and Br$\beta$ at 2.62 $\mathrm{\mu m}$) and absorption features of $\mathrm{H_2O}$ ice at 3.05 $\mathrm{\mu m}$ and $\mathrm{CO_2}$ ice at 4.27 $\mathrm{\mu m}$ are clearly detected in the spectra.
The strong recombination lines are consistent with the well-known intense starburst activity in the center of M~82, while spatially resolved $\mathrm{H_2O}$ and $\mathrm{CO_2}$ ices are another new finding although the presence of the $\mathrm{H_2O}$ ice itself was already reported by ISO (\citealt{sturm00}).
The latter will be summarized in a separate paper by using all the AKARI spectra of the ices in nearby galaxies.

It appears in Fig.2 that, as compared with the aromatic 3.3 $\mathrm{\mu m}$ emission, the aliphatic 3.4 -- 3.6 $\mathrm{\mu m}$ emission is stronger in the halo than in the center and disk. Hence we calculate the fractions of the aliphatic features to the total excess (= aromatic + aliphatic features).
In the calculation, the contribution of the aliphatic features is determined by subtracting the Drude profile, which represents the contribution of the aromatic 3.3 $\mathrm{\mu m}$ emission (\citealt{smith07}), from the total excess flux above the linear continuum between 3.3 -- 3.6 $\mathrm{\mu m}$.
We use the wavelength range of 3.20 -- 3.35 $\mathrm{\mu m}$ to fit the Drude profile; the central wavelength of the Drude profile is fixed at 3.295 $\mathrm{\mu m}$.
We first fitted the profile to the spectrum of region A3 at the galactic center and then adopted the best-fit width to the other spectra.
The best-fit Drude profile to each spectrum is shown in the red curves in Fig. 2.
We consider the errors given by the official pipeline plus the systematic errors originating in the baseline continuum; the latter errors are evaluated by changing the continuum fitting regions with shifts of $\pm$ 0.05 $\mathrm{\mu m}$.

Figure 3a shows the ratio of the aliphatic features to the total excess flux plotted against the distance from the galactic center.
We measure the distance from the peak in the 7 $\mathrm{\mu m}$ emission to the center of each sub-aperture.
As can be seen in the figure, there is a significant correlation between the ratio and distance with a linear correlation coefficient of $R=+0.75$; the ratio increases with the distance from the galactic center.
Hence Fig. 3a reveals that the strength of the aliphatic features relative to the aromatic feature is getting higher in the halo of M~82.
The colors denote north or south positions with respect to the galactic disk, i.e. the major axis of M~82 with the position angle of 64$\degr$ (\citealt{mayya05}).
In Fig. 3a, the ratios in the northern halo are higher than those in the southern halo, while there is no such systematic difference between the north and south near the center. 

\begin{figure}
\centering
\resizebox{\hsize}{!}{\includegraphics{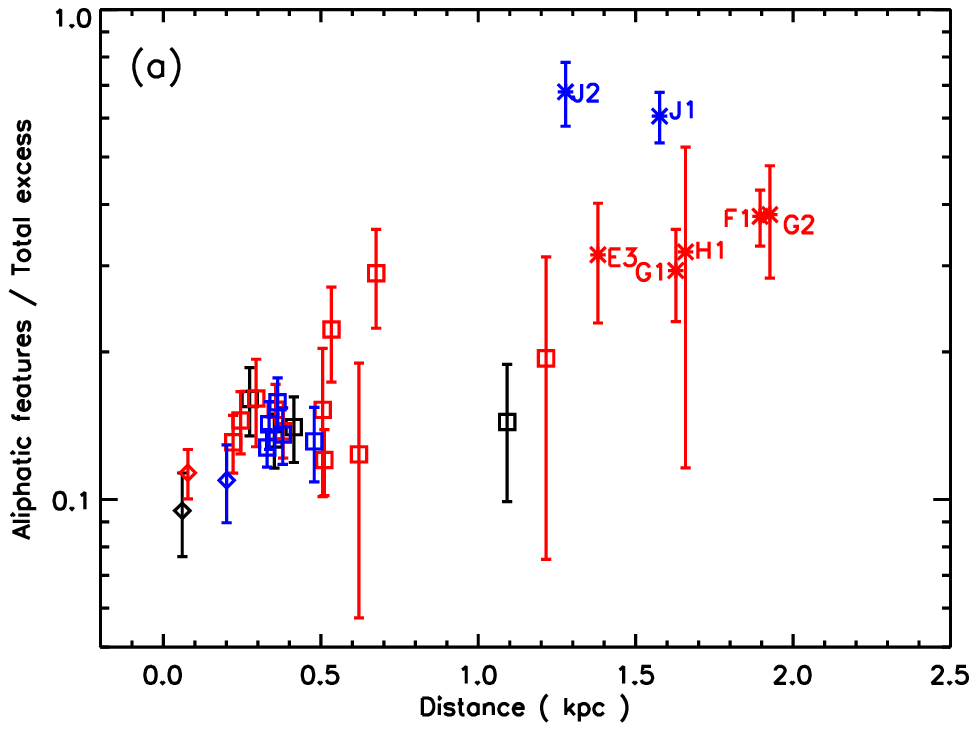}}
\resizebox{\hsize}{!}{\includegraphics{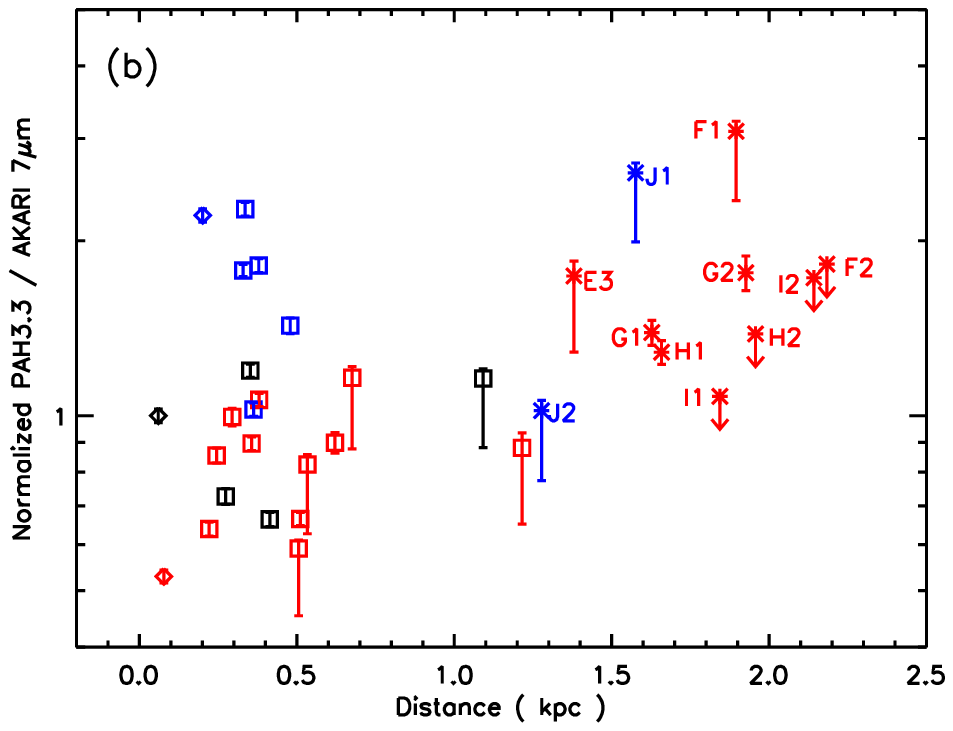}}
\caption{a) Ratio of the aliphatic features to the total excess ( = aromatic + aliphatic features) flux plotted against the distance from the center of the galaxy. The data points in the halo are labeled with the region names. The diamonds, squares, and asterisks represent the regions in the center, disk, and halo, respectively. The colors of the symbols are changed depending on the location; the black for the on-plane (A3, B3, C3-4, and E1), blue for the north side (A1-2, B1-2, C1-2, and J1-2), and red for the south side from the galactic plane (A4-6, B4-6, C5-6, D1-3, E2-3, F1-2, G1-2, H1-2, and I1-2). b) Ratio of the aromatic 3.3 $\mathrm{\mu m}$ feature intensity to the AKARI 7 $\mathrm{\mu m}$ surface brightness (\citealt{kaneda10}) plotted against the distance from the center of the galaxy. The ratios are normalized to unity in the galactic center of region A3. The definition of the symbols and colors is the same as Fig. 3a. The data points with non-detection of the PAH 3.3 $\mathrm{\mu m}$  emission are shown with the 3$\sigma$ upper limits.}
\label{fig3}
\end{figure}


\section{Discussion}

In general, the 3.3 $\mathrm{\mu m}$ feature is sensitive to PAHs of small sizes in comparison to the features at longer wavelengths (\citealt{schutte93}). Our result on the detection of the 3.3 $\mathrm{\mu m}$ feature from the halo spectra unambiguously confirms the presence of very small PAHs in the galactic superwind of M~82. By AKARI near- to mid-IR spectroscopy, \citet{Onaka10} also detected the PAH features including the 3.3 $\mathrm{\mu m}$ feature in the filamentary outflow of the starburst dwarf galaxy NGC~1569. Thus, to our knowledge, this is the second clear detection of the PAH 3.3 $\mathrm{\mu m}$ feature in galactic superwind far outside a galactic disk. \citet{Onaka10} suggested that the PAHs in the filament of NGC~1569 might be produced from the fragmentation of large carbonaceous grains in shocks. Our halo spectra in Fig.2 exhibit unusually high ratios of the aliphatic 3.4 -- 3.6 $\mathrm{\mu m}$ features to the aromatic 3.3 $\mathrm{\mu m}$ feature, supporting the 
 picture that hydrocarbon grains are highly processed in the halo. 


We compare the distribution of the PAH 3.3 $\mathrm{\mu m}$ feature with that of the PAHs features in the AKARI 7 $\mathrm{\mu m}$ band. As described above, the AKARI 7 $\mathrm{\mu m}$ band includes the PAH 6.2 and 7.7 $\mathrm{\mu m}$ features.
Figure 3b displays the ratio of the 3.3 $\mathrm{\mu m}$ feature to the AKARI 7 $\mathrm{\mu m}$ band brightness plotted against the distance from the galactic center. Here we consider that the 3.3 $\mathrm{\mu m}$ feature intensities from the spectra taken in the instrumental calibration time could be systematically overestimated by 20 \% (\citealt{mori11}), and therefore we add the 20 \% errors to the lower sides of the corresponding errorbars. Considering that the surface brightness in the 7 $\mathrm{\mu m}$ band decreases by three orders of magnitude from the center to the halo, the ratios are strikingly constant over such a wide brightness range. The result therefore indicates that the unusual hydrocarbon spectra at 3.3 -- 3.6 $\mathrm{\mu m}$ are caused by unusually strong aliphatic 3.4 -- 3.6 $\mathrm{\mu m}$ features, but not by an unusually faint aromatic 3.3 $\mathrm{\mu m}$ feature. The systematic difference between the northern and southern disk regions at $<$ 0.5 kpc from the center is most probably due to an extinction effect; it is known that M~82 is highly inclined at $\sim$ $80\degr$ and the north side of the galaxy is close to us (\citealt{lynds63}; \citealt{shopbell98}). Thus the southern disk regions are more affected by the interstellar extinction than the northern disk regions. Considering the extinction effect, the ratios tend to be more constant throughout the galaxy.


The halo spectra in Fig.2, especially those in regions J1 and J2, show remarkable resemblance to the spectra obtained at the molecular loop near the center of our Galaxy where the CO radio observations showed that the violent motion and the shock heating of gas took place (\citealt{fukui06}). \citet{kaneda11} found that the loop spectra have unusual properties at 3.3 -- 3.6 $\mathrm{\mu m}$ similarly to our halo spectra, i.e., faint 3.3 $\mathrm{\mu m}$ emission and broad excess above the continuum, from which they suggested that the destruction of small PAHs and the shattering of carbonaceous grains at the foot point of the molecular loop. In M~82, \citet{kaneda10} and \citet{roussel10} already reported the far-IR emission widely extended in the halo, showing the presence of large cold grains entrained in the galactic superwind. Hence it is possible that the unusually strong aliphatic 3.4 -- 3.6 $\mathrm{\mu m}$ features in the halo are caused by fragmentation of larger grains into small carbonaceous ones. At the same time, since we do detect the PAH features in the halo, a significant amount of PAHs including very small ones must be protected in dense molecular clouds against shocks in the halo. Alternatively, \citet{kwok11} pointed out that mixed aromatic-aliphatic organic nanoparticles can be carriers of both 3.3 and 3.4 -- 3.6 $\mathrm{\mu m}$ features. In this case, the observed unusual features may reflect simply the processing of such nanoparticles.

As can be seen in Fig.3a, the spectra in the northern halo (J1 and J2) show larger ratios of the aliphatic to the aromatic emission than those in the southern halo (E3, F1, G1-2, and H1). This result may suggest that the hydrocarbon grains are exposed to different interstellar conditions between the northern and southern halos. Comparing the spatial distribution of the $\mathrm{H_{2}}$ $v$=1-0 S(1) emission at 2.12 $\mathrm{\mu m}$ with that of the PAH emission, \citet{veilleux09} pointed out that the dominant excitation mechanism of molecular hydrogen may be different between the northern and southern halos. They suggested that molecular hydrogen may be excited mostly by UV radiation in the southern halo because they have similar distributions there; it is also expected that photoionization by OB stars is dominant in the southern halo region (\citealt{shopbell98}). In contrast, their distributions are significantly different in the northern halo (\citealt{veilleux09}). Our result in Fig.3a favors the scenario that shocks are more important for the heating of molecular hydrogen in the northern halo where hydrocarbon grains are more efficiently processed and destroyed.

\section{Conclusions}

With AKARI, we have performed near-IR (2.5 -- 4.5 $\mathrm{\mu m}$) spectroscopic observations of the nearby edge-on starburst galaxy M~82. We obtain 34 spectra in the galaxy including the northern and southern halos to investigate the properties of carbonaceous grains in the galactic superwind.
In many spectra, aromatic 3.3 $\mathrm{\mu m}$ emission and aliphatic 3.4 -- 3.6 $\mathrm{\mu m}$ emission are clearly detected.
In particular, the aromatic and aliphatic emissions are detected in halo regions 2 kpc far away from the galactic center, which clearly demonstrates that even small PAHs survive in a harsh environment of the galactic superwind.
We find that the fraction of the aliphatic to the aromatic feature increases with the distance from the galactic center. At the same time, we show that the strength of the PAH 3.3 $\mathrm{\mu m}$ feature relative to that of the PAH 6.2 and 7.7 $\mathrm{\mu m}$ features does not decrease toward the halo.
Therefore we conclude that the unusual hydrocarbon spectra at 3.3 -- 3.6 $\mathrm{\mu m}$ are caused by unusually strong aliphatic 3.4 -- 3.6 $\mathrm{\mu m}$ features, but not by an unusually faint aromatic 3.3 $\mathrm{\mu m}$ feature. The result suggests that, in the galactic superwind, small carbonaceous grains emitting the aliphatic features are produced by fragmentation of larger grains in shocks, while an appreciable amount of small PAHs are protected in dense molecular clouds against shocks.

\begin{acknowledgements}
We thank all the members of the AKARI project.
We also express many thanks to the anonymous referee for the useful comments.
AKARI is a JAXA project with the participation of ESA.
This work is supported by a Grant-in-Aid for Japan Society for the Promotion of Science Fellows, 23005457.
\end{acknowledgements}


\end{document}